# Experimental demonstration of quantum memory for light

Brian Julsgaard[1], Jacob Sherson[1,2], J. Ignacio Cirac[3], Jaromír Fiurášek[4], and Eugene S. Polzik[1]

[1] *Niels Bohr Institute, Danish Quantum Optics Center – QUANTOP, Copenhagen University, Blegdamsvej 17, 2100 Copenhagen Ø, Denmark.*

[2] *Department of physics, Danish Quantum Optics Center – QUANTOP, University of Aarhus, 8000 Aarhus C, Denmark.*

[3] *Max Planck Institute for Quantum Optics, Hans-Kopfermann-Str. 1, D-85748 Garching, Germany.*

[4] QUIC, Ecole Polytechnique, CP 165, Universite Libre de Bruxelles, *1050 Brussels, Belgium and Department of Optics, Palacky University, 17. listopadu 50, 77200 Olomouc, Czech Republic*

**The information carrier of today's communications, a weak pulse of light, is an intrinsically quantum object. As a consequence, complete information about the pulse cannot, even in principle, be perfectly recorded in a classical memory. In the field of quantum information this has led to a long standing challenge: how to achieve a high-fidelity transfer of an independently prepared quantum state of light onto the atomic quantum state[1-4]? Here we propose and experimentally demonstrate a protocol for such quantum memory based on atomic ensembles. We demonstrate for the first time a recording of an externally provided quantum state of light onto the atomic quantum memory with a fidelity up to 70%, significantly**

**higher than that for the classical recording. Quantum storage of light is achieved in three steps: an interaction of light with atoms, the subsequent measurement on the transmitted light, and the feedback onto the atoms conditioned on the measurement result. Density of recorded states 33% higher than that for the best classical recording of light on atoms is achieved. A quantum memory lifetime of up to 4 msec is demonstrated.**

Light is a natural carrier of information in both classical and quantum communications. In classical communications, bits are encoded in large average amplitudes of light pulses which are detected, converted into electric signals, and subsequently stored as charges or magnetization of memory cells. In quantum information processing, information is encoded in quantum states that cannot be accurately recorded by such classical means. Consider a state of light defined by its amplitude and phase, or equivalently by two quadrature phase operators, $\hat{X}_L$ and $\hat{P}_L$, with the canonical commutation relation $\left[\hat{X}_L, \hat{P}_L\right] = i$. These variables play the same role in quantum mechanics as the classical quadratures $X,P$ do in the decomposition of the electric field of light with the frequency $\omega$ as $E \propto X \cos \omega t + P \sin \omega t$. Other quantum properties of light, such as the photon number $\hat{n} = \frac{1}{2}\left(\hat{X}_L^2 + \hat{P}_L^2 - 1\right)$, etc., can be expressed in terms of $\hat{X}_L$ and $\hat{P}_L$.

The best classical approach to recording a state of light onto atoms would involve homodyne *measurements of both* observables $\hat{X}_L$ and $\hat{P}_L$ by using, e.g., a beam splitter. The non-commutativity of $\hat{X}_L$ and $\hat{P}_L$ leads to additional quantum noise added during this procedure. The target atomic state has its intrinsic quantum noise (coming from the Heisenberg uncertainty relations). All this extra noise leads to a limited fidelity for the classical recording, e.g., to a maximum fidelity of 50% for coherent states[5-7]. Thus the challenge of implementing a quantum memory can be formulated as a faithful storing of the simultaneously immeasurable values of $\hat{X}_L$ and $\hat{P}_L$.

A number of quantum information protocols, such as eavesdropping in quantum cryptography, quantum repeaters[8], and linear optics quantum computing[9] would benefit from a memory meeting the following criteria: 1.The light pulse to be stored is sent by a third party in a state unknown to the memory party. 2. The state of light is converted into a quantum state of the memory with a fidelity higher than that of the classical recording. Several recent experiments[10-13] have demonstrated entanglement of light and atoms. However, none of these experiments demonstrated the memory obeying the two above criteria. In ref. 14, where squeezed light was mapped onto atoms, the atomic state existed only while the light was on, so it was not a memory device. The electromagnetically induced transparency (EIT) approach has led to the demonstration of a classical memory for light[15,16]. A theoretical proposal for EIT-based quantum memory for light has been published in ref. 3. Other proposals for quantum memory for light with better-than-classical quality of recording have also been published recently[1-4].

Quantum state transfer from one species to another is most simply presented if both systems are described by canonical quantum variables $\hat{X}, \hat{P}$. All canonical variables have the same commutation relations and the same quantum noise for a given state, providing thus a common frame for the analysis of the state transfer.

In the present work the state of light is stored in the superposition of magnetic sublevels of the ground state of an atomic ensemble. As in ref. 12, we introduce the operator $\hat{J}$ of the collective magnetic moment (orientation) of a ground state $F$. All atomic states utilized here are not too far in phase space from the coherent spin state (CSS) for which only one projection has a non-zero mean value, e.g., $\langle \hat{J}_x \rangle = J_x$ whereas the other two projection have minimal quantum uncertainties, $\langle \delta J_y^2 \rangle = \langle \delta J_z^2 \rangle = \frac{1}{2} J_x$. For all such states the commutator $\left[ \hat{J}_y, \hat{J}_z \right] = i J_x$ can be reduced to the canonical commutator $\left[ \hat{X}_A, \hat{P}_A \right] = i$ with $\hat{X}_A = \hat{J}_y / \sqrt{J_x}, \hat{P}_A = \hat{J}_z / \sqrt{J_x}$. Hence the $y,z$-components of the collective atomic angular momentum play the role of canonical variables. Although the memory protocol, in principle, can work with a single atomic ensemble, experimental technical noise is substantially reduced if two oppositely polarized ensembles placed in a bias magnetic field $\vec{H}$ are used (see Methods sections and Supplementary Methods for details). Combined canonical variables for two ensembles $\hat{X}_A = (\hat{J}_{y1} - \hat{J}_{y2}) / \sqrt{2J_x}$, $\hat{P}_A = (\hat{J}_{z1} + \hat{J}_{z2}) / \sqrt{2J_x}$ are then introduced where $\hat{J}_{x1} = -\hat{J}_{x2} = J_x = FN_{atoms}$. In the presence of $\mathbf{H}$ the memory couples to the $\Omega$-sidebands of light: $\hat{X}_L = \frac{1}{\sqrt{T}} \int_0^T (\hat{a}^+(t) + \hat{a}(t)) \cos(\Omega t) dt$, $\hat{P}_L = \frac{i}{\sqrt{T}} \int_0^T (\hat{a}^+(t) - \hat{a}(t)) \cos(\Omega t) dt$, where $\Omega$ is the Larmor frequency of spin precession.

Quantum storage of light is achieved in three steps: (1) an interaction of light with atoms; (2) a subsequent measurement of the transmitted light; and (3) Feedback onto the atoms conditioned on the measurement result (Fig. 1). The off-resonant interaction of light with spin polarized atomic ensembles has been described elsewhere[4,17-19], and is summarized in the Methods section. The interaction leads to the equations

$$\hat{X}_L^{out} = \hat{X}_L^{in} + k\hat{P}_A^{in}, \hat{P}_L^{out} = \hat{P}_L^{in} \qquad (1)$$
$$\hat{X}_A^{out} = \hat{X}_A^{in} + k\hat{P}_L^{in}, \hat{P}_A^{out} = \hat{P}_A^{in}$$

These equations imply that light and atoms get entangled. The remarkable simplicity of equations (1) provides a direct link between an input light state, an atomic state, and the output light. Suppose the input light is in a vacuum (or in a coherent) state and atoms are in a CSS with mean values $\langle \hat{X}_L \rangle = \langle \hat{X}_A \rangle = \langle \hat{P}_L \rangle = \langle \hat{P}_A \rangle = 0$ and variances $\delta X_L^2 = \delta X_A^2 = \delta P_L^2 = \delta P_A^2 = \frac{1}{2}$. The interaction parameter $k$ whose value is crucial for the storage protocol is then readily found as $k^2 = 2(\delta X_L^{out})^2 - 1$.

For a perfect fidelity of mapping, the initial atomic state must be an entangled spin state such as in ref. 12, with $\delta X_A^2 \rightarrow 0$. The pulse to be recorded, combined with the entangling pulse (see Methods section), is sent through, and its variable $\hat{X}_L^{out}$ is measured. The measurement outcome, $x = \hat{X}_L^{in} + k\hat{P}_A^{in}$, is fed back into the atomic variable $\hat{P}_A$ with a feedback gain $g$. The result is $\hat{P}_A^{mem} = \hat{P}_A^{in} - gx = \hat{P}_A^{in}(1-kg) - g\hat{X}_L^{in}$ (see Supplementary Notes for justification of this equation). With $g = k = 1$, the mapping of $\hat{X}_L^{in}$ onto $-\hat{P}_A^{mem}$ is perfect.

The second operator of light is already mapped onto atoms via $\hat{X}_A^{mem} = \hat{X}_A^{in} + \hat{P}_L^{in}$ see equation (1). For the entangled initial state the mapping is perfect for this component too, $\hat{P}_L^{in} \rightarrow \hat{X}_A^{mem}$, leading to the fidelity of the light-to-atoms state transfer $F \rightarrow 100\%$. If the initial atomic state is a CSS the mapping is not perfect due to the noisy operator $\hat{X}_A^{in}$. However, $F = 82\%$, still markedly higher than the classical limit, can be achieved. Note that the above discussion holds for an arbitrary single mode input quantum state of light.

In our experiment the atomic storage unit consists of two samples of Cesium vapor placed in paraffin coated glass cells placed inside magnetic shields (Fig. 1). **H** is applied along the $x$-direction with $\Omega = 322$kHz. Optical pumping along **H** initializes the atoms in the first/second sample in the $F = 4, m_F = \pm 4$ ground state with the orientation above 99%. Hence $\hat{J}_{x1} = -\hat{J}_{x2} = J_x = 4N_{atoms} \approx 1.2 \times 10^{12}$. We thoroughly check and regularly verify that the initial spin state is close to CSS (Supplementary Methods). The coupling parameter $k$ is varied by adjusting the density of Cs.

The input state $\hat{a}(t)$ is encoded in a 1-msec $y$-polarized pulse. The state is chosen from the set $\{\hat{a}_{input}\}$ of coherent states with the photon number in the range $\left\{ \langle n \rangle = 0, n_{max} \right\}$ and an arbitrary phase. $\hat{a}(t)$ is generated as $\Omega$-sidebands by an electro-optical modulator (EOM) and has the same spatial and temporal profile as the strong entangling field (more information can be found in the Methods section). Thus the EOM plays the third party, providing the field to be stored. The pulses are detuned by 700 MHz to the blue from the $6S_{1/2}$, $F$=4 ? $6P_{3/2}$, $F$=5 transition ($\lambda$=852nm). The polarization measurement of the light is followed by the feedback onto atoms achieved by a 0.2ms radio-frequency magnetic pulse conditioned on the measurement result.

Next the experimental verification of the quantum storage is carried out. A read-out *x*-polarized pulse is sent through the samples with the delay of 0.7-10 milliseconds after the feedback is applied. Atomic memory generates a *y*-polarized pulse which is analyzed as follows. Since both $\hat{X}_A^{mem}$ and $\hat{P}_A^{mem}$ cannot be measured at the same time, we carry out two series of measurements for each input state. Each series consists of $10^4$ quantum storage sequences. To verify the $\hat{X}_L^{in} \rightarrow -\hat{P}_A^{mem}$ step of the storage, we measure the component $\hat{X}_L^{read-out} = \hat{X}_L^{read-in} + k\hat{P}_A^{mem}$ of the read-out pulse ($X_L$ is a Stokes parameter measured in units of shot noise as discussed in the Methods section). An example of such a measurement carried out after 0.7 msec of storage is presented in Fig. 2a as a histogram of $\frac{1}{k}\hat{X}_L^{read-out}$ (right histogram) with *k* measured as described in the Methods section and in Supplementary methods. For this series $\langle \hat{P}_L^{in} \rangle = -4$ and $\langle \hat{X}_L^{in} \rangle = 0$ corresponding to $\langle \hat{n} \rangle = 8$ photons in the pulse. From this measurement we find the mean $\langle \hat{P}_A^{mem} \rangle = \frac{1}{k}\langle X_L^{read-out} \rangle$ and the variance $\sigma_p^2 = \left\langle \left( \delta \hat{P}_A^{mem} \right)^2 \right\rangle = \frac{1}{k^2}\left( \left( \delta X_L^{read-out} \right)^2 - \frac{1}{2} \right)$ (see equation 1) for the quantum state of the memory. We note that only the knowledge of *k* and the shot noise level of light is necessary for the determination of the mean values and variances of the atomic canonical variables from the experimental data.

Next we run another series of storage with the same input state for the verification of the step $\hat{P}_L^{in} \rightarrow \hat{X}_A^{mem}$. The $\hat{X}_A^{mem}$ operator does not couple to the read-out pulse in our geometry; therefore, we first apply a $\pi/2$-pulse (Fig. 1) to atoms converting $\hat{X}_A^{mem} \rightarrow \hat{P}_A^p$ and then measure $\hat{P}_A^p$ with the verifying pulse. We then find $\langle \hat{X}_A^{mem} \rangle$ and $\sigma_x^2 = \left\langle \left( \delta \hat{X}_A^{mem} \right)^2 \right\rangle$ of the memory state (left histogram).

The above sequence is repeated for different input states. From $\langle \hat{P}_A^{mem} \rangle / \langle \hat{X}_L^{in} \rangle$ and $\langle \hat{X}_A^{mem} \rangle / \langle \hat{P}_L^{in} \rangle$ the mapping gains for the two quadratures are determined. For the experimental data presented in Fig.2 and 3a, these gains are 0.80 and 0.84 respectively, which is close to the optimal gain for the chosen input set of states. This step would complete the proof of the *classical* memory performance, because we have shown that the *y*-polarized pulse recovered from the memory has the same *mean* amplitude and *mean* phase as the input pulse (up to a chosen constant factor).

To prove a quantum memory performance we need in addition to consider the quantum noise of the stored state. Towards this end we plot the atomic variances $\sigma_p^2, \sigma_x^2$ for the storage time 0.7 msec in Fig. 3a. The experimentally obtained variances of the stored state are on average 33% below the best possible variance of the classical recording. Hence the density of stored states 33% higher than that for the best classical recording can be obtained. Thus the goal of quantum storage with less noise than for the classical recording is achieved.

Next the overlap between the input state of light and the state of the atomic memory is determined (Methods section). An example is shown in Fig. 2b. The fidelity F of the quantum recording is then calculated for a given set $\{\hat{a}_{\text{input}}\}$. For example, $F=(66.7\pm1.7)\%$ for $\{\hat{a}_{\text{input}}\}=\{n=0\rightarrow8\}$ and $F=(70.0\pm2.0)\%$ for $\{\hat{a}_{\text{input}}\}=\{n=0\rightarrow4\}$, respectively for the storage time of 0.7 msec. Note that the fidelity of the classical recording can exceed 50% for a limited set $\{\hat{a}_{\text{input}}\}$. The maximum classical fidelity for $\{\hat{a}_{\text{input}}\}=\{n=0\rightarrow8\}$ is 55.4%, and for $\{\hat{a}_{\text{input}}\}=\{n=0\rightarrow4\}$ it is 59.6% - still significantly lower than that for the quantum recording.

The main sources of imperfection of our quantum memory are decoherence of the atomic state and reflection off the cell walls. We have performed extensive studies of the atomic decoherence caused by the light-assisted collisional relaxation[20] to optimize the fidelity. Fig. 3b presents the fidelity of the stored state as a function of the storage time. A simple model provides a good description for the observed fidelity reduction.

The single observable read-out described above can be useful, e.g., in quantum cryptography eavesdropping, where the memory is read after the basis has been publicly announced by Alice and Bob. The present experiment also paves the road towards the proposed quantum cloning of light onto atomic memory[21]. However, other applications require complete state recovery via reverse mapping of the memory state onto light. Proposals for performing this task within our approach have been published[4,19,22]. Probably the most intuitively clear protocol for the memory read-out is just to run the storage protocol of the present paper with the reversed roles of light and atoms. Indeed the equations of interaction (1) are completely symmetric. The read-out, as the storage, would involve three steps: sending a read-out light pulse through atoms, measuring the spin projection $\hat{X}_A^{out}$ with an auxiliary light pulse, and applying the feedback conditioned on this measurement to the read-out pulse.

In the present experiment we have demonstrated the memory for a subset of linearly independent coherent states. Due to the linearity of quantum mechanics this demonstration signifies that our method provides faithful mapping for an arbitrary coherent state. Since any arbitrary quantum state can be written as a superposition of coherent states, our approach should in principle work for an arbitrary quantum state, including entangled and single photon (qubit) states.

**Methods**

**Quantum coupling of light to two atomic ensembles in the presence of magnetic field**

Here we discuss the physics behind the equations of interaction (1). The off-resonant atom/light interaction is described in terms of Stokes operators for the polarization state of light and the collective spin of atoms[4,17,18]. The Stokes operators are defined as one half of the photon number difference between orthogonal polarization modes: $\hat{S}_1$- between vertical $x$- and horizontal $y$-polarizations, $\hat{S}_2$ - between the modes polarized at $\pm 45^0$ to the vertical axis, and $\hat{S}_3$ - between the left- and right-hand circular polarizations. In the experiment a strong entangling $x$-polarized pulse with the photon flux $n(t)$ is mixed on a polarizing beamsplitter with the $y$-polarized quantum field $\hat{a}(t)$ prior to interaction with atoms. Hence the Stokes operators of the total optical field are $\hat{S}_1(t) = S_1(t) = \frac{1}{2}n(t)$, $\hat{S}_2 = \frac{1}{2}\sqrt{n(t)}(\hat{a}^+(t) + \hat{a}(t))$, $\hat{S}_3 = \frac{i}{2}\sqrt{n(t)}(\hat{a}^+(t) - \hat{a}(t))$. Note that $\hat{S}_2(t)$ and $\hat{S}_3(t)$ are proportional to the canonical variables for the quantum light mode $\hat{X} = \frac{1}{\sqrt{2}}(\hat{a}^+(t) + \hat{a}(t)), \hat{P} = \frac{i}{\sqrt{2}}(\hat{a}^+(t) - \hat{a}(t))$. Light is transmitted through the atomic samples placed in the bias magnetic field oriented along the $x$-axis. The magnetic field allows for encoding of the memory at the Larmor frequency $\Omega$, thus dramatically reducing technical noise present at low frequencies. However, in the presence of the Larmor precession, there is an undesired coupling of the single cell variables $\hat{J}_y$ and $\hat{J}_z$ to each other. The introduction of the second cell with the opposite Larmor precession allows us to introduce new two-cell variables $(\hat{J}_{y1} - \hat{J}_{y2}), (\hat{J}_{z1} + \hat{J}_{z2})$ that do not couple to each other. As in ref. 12, where a similar trick was used, the Stokes parameters of light transmitted through the two cells along the $z$ direction become

$$\hat{S}_2^{out}(t) = \hat{S}_2^{in}(t) + aS_1\Big(\cos(\Omega t)[\hat{J}_{z1} + \hat{J}_{z2}] + \sin(\Omega t)[\hat{J}_{y1} + \hat{J}_{y2}]\Big), \hat{S}_3^{out}(t) = \hat{S}_3^{in}(t) \qquad (2)$$

where $\hat{J}_{z,y}$ are the projections in the frame rotating at $\Omega$ and $a = \dfrac{\gamma\lambda^2}{8\pi\Delta A}$, with $\gamma$ and $\lambda$ - the natural linewidth and the wavelength of the transition respectively, $\Delta$ - the detuning, and $A$ - the beam cross-section. At the same time, the transverse spin components of the two cells evolve as follows:

$$\frac{d}{dt}[\hat{J}_{z1} + \hat{J}_{z2}] = \frac{d}{dt}[\hat{J}_{y1} + \hat{J}_{y2}] = 0,$$
$$\frac{d}{dt}[\hat{J}_{y1} - \hat{J}_{y2}] = 2aJ_x\hat{S}_3^{in}\cos(\Omega t), \frac{d}{dt}[\hat{J}_{z1} - \hat{J}_{z2}] = 2aJ_x\hat{S}_3^{in}\sin(\Omega t)$$

$$(3)$$

As evident from equation (3), in the process of propagation the operator $\hat{S}_3^{\text{in}}$ is recorded onto the operators $\hat{J}_{y1} - \hat{J}_{y2}$ and $\hat{J}_{z1} - \hat{J}_{z2}$ (the "back action" of light on atoms via the dynamic Stark effect caused by light[17,18]), while the operators $\hat{J}_{y1} + \hat{J}_{y2}$ and $\hat{J}_{z1} + \hat{J}_{z2}$ are left unchanged. The latter are read out onto $\hat{S}_2^{\text{out}}$ via the Faraday rotation (2).

Canonical variables are defined for the quantum light mode as $\hat{X}_L = \frac{1}{\sqrt{T}} \int\limits_0^T (\hat{a}^+(t) + \hat{a}(t))\cos(\Omega t)dt$, $\hat{P}_L = \frac{i}{\sqrt{T}} \int\limits_0^T (\hat{a}^+(t) - \hat{a}(t))\cos(\Omega t)dt$, that is the relevant light mode involves the O-sidebands. $T$ is the pulse duration, $\hat{a}(t)$ is normalized to the photon flux. $\hat{X}_L$ and $\hat{P}_L$ (i.e., $\hat{S}_2$ and $\hat{S}_3$) are detected by a polarization state analyzer and by lock-in detection of the O component of the photocurrent. Note that the $\cos(\Omega t)$ component of light couples to the $(\hat{J}_{y1} - \hat{J}_{y2}), (\hat{J}_{z1} + \hat{J}_{z2})$ components of atomic storage variables (equations (2,3)). The equivalent choice of a $\sin(\Omega t)$ modulation instead would mean the use of $(\hat{J}_{y1} + \hat{J}_{y2}), (\hat{J}_{z1} - \hat{J}_{z2})$ for the memory. The atomic canonical variables $\hat{X}_A, \hat{P}_A$ are defined in the main section. With the above equations and definitions we straightforwardly derive equation (1) under the assumption $\Omega T \gg 1$. Theoretically the dimensionless coupling parameter in equation (1) is $k^2 = \frac{1}{2} a^2 J_x \int n(t)dt$.

**Experimental calibration of the canonical variances for light and atoms**

Calculations of the fidelity, the gains, and the variances from the experimental data are based on the experimental calibration of $\left\langle \delta \hat{X}_L^2 \right\rangle = \left\langle \delta \hat{P}_L^2 \right\rangle$ for the coherent (vacuum) state of light and of $\left\langle \delta \hat{X}_A^2 \right\rangle = \left\langle \delta \hat{P}_A^2 \right\rangle$ for the coherent spin state (CSS) of atoms. The calibration for light is carried out along the established procedure of determining the shot noise level for measurements of $\hat{S}_2, \hat{S}_3$ with the quantum field in a vacuum state[5,17]. Variances and mean values for light are then measured in units of this shot noise level. The calibration for the atomic CSS variance is carried out with extreme care and has shown excellent reproducibility (See Supplementary Methods). As stated in the main text, as soon as the vacuum (shot) noise level for light is established and the atoms are in a CSS, the parameter $k^2$ (equation 1), important for calculations of atomic variances and fidelity, is easily determined as $k^2 = 2\left(\delta X_L^{out}\right)^2 - 1$. In the experiment this is equivalent to $k^2 = \left(\left(\delta S_2^{out}\right)^2 - \left(\delta S_2^{in}\right)^2\right) / \left(\delta S_2^{in}\right)^2$.

**Fidelity and the state overlap**

To calculate the fidelity of the transfer of an input coherent state into an output Gaussian state[6], we first define an overlap function between an input state with mean values $x_1, p_1$ and the output state with the mean values and variances $x_2, p_2, \sigma_x^2, \sigma_p^2$.

Straightforward integration yields

$O\{x_1, x_2, p_1, p_2\} = 2\exp\left(- (x_1 - x_2)^2 / (1 + 2\sigma_x^2) - (p_1 - p_2)^2 / (1 + 2\sigma_p^2)\right) / \sqrt{(1 + 2\sigma_x^2)(1 + 2\sigma_p^2)}$

. The fidelity of the transfer for a set of coherent states with mean amplitudes between $\alpha_1$ and $\alpha_2$ can then be found as an average overlap $F = \pi^{-1}(\alpha_2^2 - \alpha_1^2)^{-1} \int_0^{2\pi} d\phi \int_{\alpha_1}^{\alpha_2} O\{\alpha\} \alpha d\alpha$.

For classical recording from light onto atoms with the gain $g$, the overlap between the input coherent state with the mean amplitude $\alpha = \sqrt{x^2 + p^2}$ and the output state is given by $O\{\alpha\} = (1 + g^2)^{-1} \exp\left(-\frac{1}{2}(1 - g)^2 \alpha^2 (1 + g^2)^{-1}\right)$. The classical fidelity is then given by $F_{class} = (n_2 - n_1)^{-1}(1 - g)^{-2} \left\{\exp\left(-(1 - g)^2 n_1 (1 + g^2)^{-1}\right) - \exp\left(-(1 - g)^2 n_2 (1 + g^2)^{-1}\right)\right\}$ where we have introduced the mean photon number $n = \frac{1}{2}\alpha^2$. $F_{class} \rightarrow 50\%$ for arbitrary coherent states when $g \rightarrow 1$. If a restricted class of coherent states is chosen as the input, $F_{class} > 50\%$ can be obtained with a suitable choice of $g$. For a set of states analyzed in the main text, $\{\hat{a}_{input}\} = \{n = 0 \rightarrow 8\}$, the maximum classical fidelity of 55.4% is achieved with the gain of 0.809.

We are grateful to Nicolas Cerf and Klemens Hammerer for illuminating discussions. This research was funded by the Danish National Research Foundation, by EU grants QUICOV and COVAQIAL, and by the project "Research Center for Optics" of the Czech Ministry of Education." IC and ESP gratefully acknowledge the hospitality of the Institute for Photonic Sciences in Barcelona where part of this work was initiated.

**Correspondence and requests for materials should be addressed to Eugene Polzik, polzik@nbi.dk**

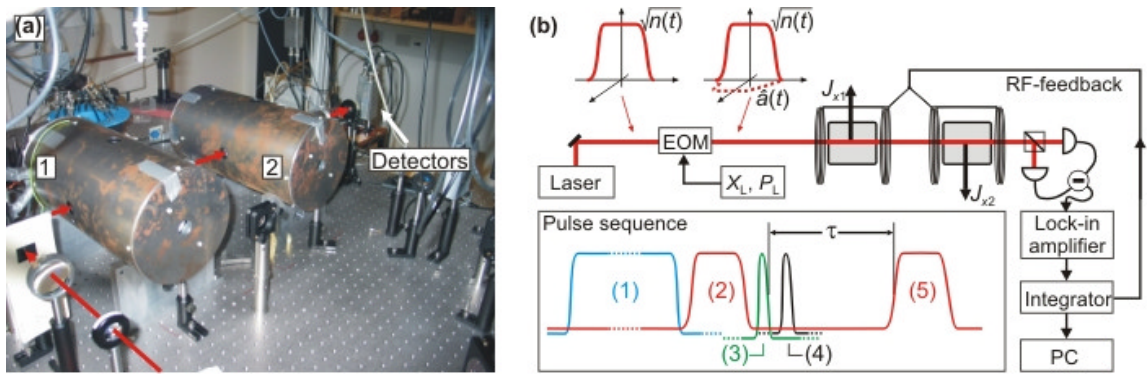

Figure 1. Experimental setup. a) Atomic memory unit consisting of two Cesium cells inside magnetic shields 1, 2. The path of the recorded and read-out light pulses is shown with arrows. b) The simplified layout of the experiment. The input state of light with the desired displacements $X_L$, $P_L$ is generated with the electro-optic modulator (EOM). The inset shows the pulse sequence for the quantum memory recording and read-out. Pulse 1 is the optical pumping (4ms), pulse 2 is the input light pulse $\hat{a}(t)$ overlapped with the strong entangling pulse in orthogonal polarization with the amplitude $\sqrt{n(t)}$. Pulse 3 is the magnetic feedback pulse. Pulse 4 is the magnetic $\pi/2$ pulse used for the read out of one of the atomic operators. Pulse 5 is the read-out optical pulse.

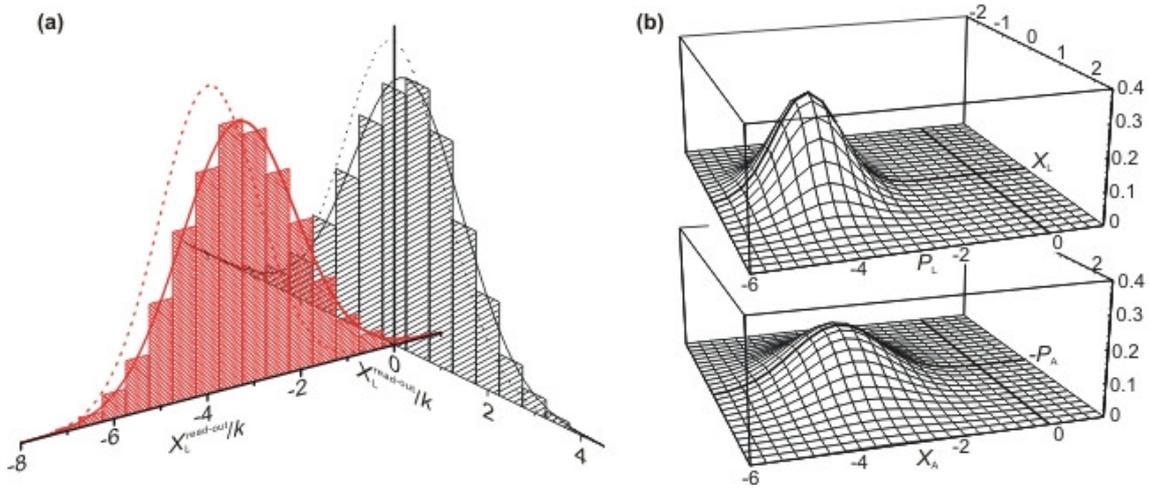

Figure 2. An example of the atomic memory performance. a). The input state of light in the coherent state with $\langle \hat{X}_L \rangle = 0, \langle \hat{P}_L \rangle = -4$. The results of the read out of this state stored in the atomic memory are shown as histograms of experimental realizations. The left/right histogram shows the results for the $\hat{X}_A / \hat{P}_A$ quadrature read out with/without the $\frac{\pi}{2}$-pulse. Dotted Gaussians represent the distributions for the best possible quantum memory performance (fidelity 100%). b). The input coherent state of light (upper graph) and the reconstructed state stored in the atomic memory (lower graph) for the input state as in figure 2a. The reconstructed state is obtained from the results presented in figure 2a after subtracting the noise of the read out pulse.

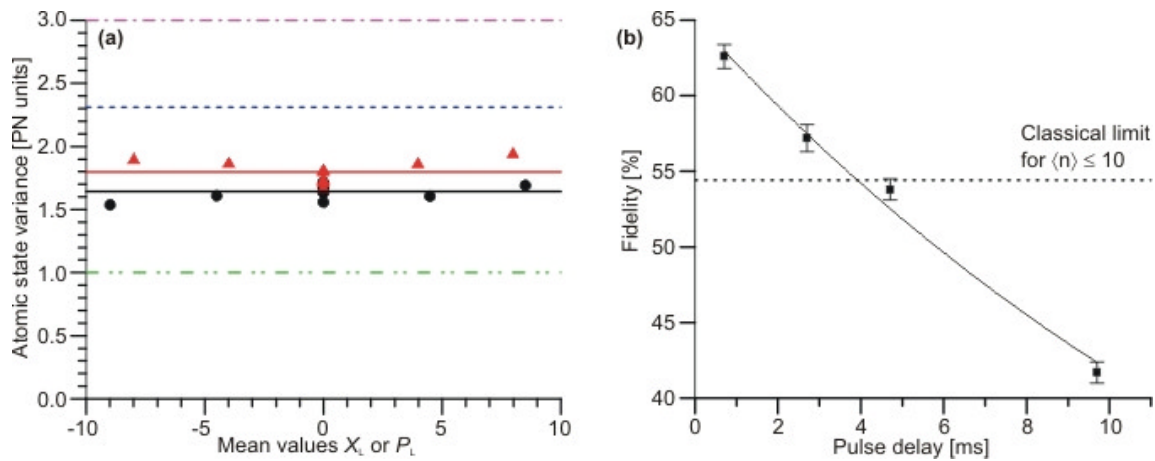

Figure 3, Quantum noise of the stored state and the fidelity of quantum memory as a function of time. a). Experimental and theoretical (quantum and classical) stored state variances in atomic projection noise (PN) units. Trianges and filled circles are the experimental variances for the atomic memory operators, denoted $2\sigma_x^2$ and $2\sigma_p^2$ respectively in the text. Dash-dotted line – the fundamental boundary of three units of noise between quantum and classical mapping for an arbitrary coherent input state[5,6]. Dashed line – best classical variance for the experimental set of input states with photon numbers between 0 and 8. Double-dot-dashed line – the unity variance corresponding to perfect mapping. b). Fidelity as a function of storage time for the set of states from 0 to 10 photons. Fidelity higher than the classical limit is maintained for up to 4 msec of storage. Error bars (std. dev.) include fitting uncertainty of gains and variances and an additional uncertainty of 2.5% in the projection noise calibration.

**Supplementary Methods**

**Calibration of the Atomic Projection Noise level**

From the measurement of the variance of the Stokes parameter $\hat{S}_2^{\text{out}}$ as a function of the macroscopic spin size $J_x$, we can determine the contribution of the atomic projection noise to the noise of the transmitted light. The goal here is to measure the light-atoms coupling parameter $k$ that is used for the calculation of the canonical atomic variables in the paper. It is convenient that we do not need to know explicitly the absolute value of the projection noise (in the sense of $\left\langle \hat{J}_{z,y}^2 \right\rangle = \frac{F}{2} N_{atoms}$ ). However, we do need to determine the projection noise contribution to the light noise.

In order to determine this contribution we need to (1) extract the linear dependence of $\left( \delta \hat{S}_2^{\text{out}} \right)^2$ on $J_x$; and (2) Ensure that the atoms are spin polarized to a high degree.

The atomic spin noise is measured for two cells together according to the combined two-cell quantum variables introduced in the main text. As any modulation technique, this approach allows to overcome technical noise by means of lock-in detection at the modulation frequency. In our case we have been able in this way to eliminate technical noise to well below the $10^{-6}$ level, and thus reach the quantum projection noise limit for up to $3*10^{11}$ atoms.

The atoms are optically pumped with a 4 msec pulse preparing a fresh state before each measurement. The Stokes parameter $\hat{S}_2 = \frac{1}{\sqrt{2T}} \int_0^T \sqrt{n(t)} (\hat{a}^+(t) + \hat{a}(t)) \cos(\Omega t) dt$ is measured by the lock-in detection. The shot noise of the incoming light $\left( \delta \hat{S}_2^{\text{in}} \right)^2$ is measured separately. Repeating the optical pumping and the measurement sequence many times, we obtain the variance of the operator $\hat{S}_2^{\text{out}}$. By measuring the Faraday rotation angle $\phi$ of a linearly polarized light propagating along the $x$ direction of the macroscopic spin polarization we obtain the value proportional to the ensemble mean spin $J_x$. We also determine the degree of optical pumping (spin orientation of the ground state $F = 4$) by the magneto-optical resonance method[20]. We routinely find a degree of optical pumping better than 99%.

In the figure (part (a)) we plot the atomic contribution to the variance of the transmitted light normalized to the shot noise level: $\left(\left(\delta S_2^{out}\right)^2 - \left(\delta S_2^{in}\right)^2\right)/\left(\delta S_2^{in}\right)^2$, as a function of $\phi$. The value of $J_x$ is varied by varying the temperature of the sample. The lower part of the graph shows a nice linear dependence (solid line) which together with a nearly perfect degree of orientation proves that we observe quantum spin noise, i.e., the projection noise of the coherent spin state (CSS) (while classical noise would grow quadratically with $J_x$). The scattering of the points, especially at high atomic densities, arises from the technical laser noise, as proven by an independent monitoring of this noise.

The above procedure has been carried out on a regular basis to ensure that the contribution of the projection noise is reliably defined. We find that, provided the geometry, detuning, duration, and power of the light beam are carefully reproduced, the excess noise of the laser controlled, and the magnetic shielding of the atoms sufficient, the PNL contribution can be determined with a high level of confidence.

As an example, in part (b) of the figure we show the PNL calibration 43 days after the data (a) was obtained. The solid line here is the same as in part (a) and it neatly coincides with a linear fit through zero of the lower half of the points. We have thus a reproducible PNL calibration.

The procedure described above is quite similar to the determination of the shot noise level of polarized light, a routine well established in the studies of squeezed and entangled light (except, of course, that atoms replace photons in our case). There, similarly to the present work, as soon as light is well polarized, the linear dependence of the noise variance on the photon number (power) signifies that the coherent state noise (shot noise) level is achieved.

The PNL is estimated to be stable to within 2.5% and this number is used in the text to calculate the uncertainty of the fidelity $F = (66.7 \pm 1.6)\%$. However, the PNL uncertainty plays only a minor role here. For example, with a 10% uncertainty in PNL we would get $F = (66.7 \pm 2.6)\%$. The reason for the weak dependence of the fidelity uncertainty on the PNL uncertainty can be understood as follows: if the PNL is higher than estimated, the variance of the stored state is actually lower (in the PNL units) which leads to a higher fidelity. But at the same time the gain factor is also lower leading to a lower fidelity. The two effects oppose each other and hence the fidelity is a rather slowly varying function of PNL.

The parameter $k^2$ is determined from the linear contribution to the function $\left(\left(\delta S_2^{out}\right)^2 - \left(\delta S_2^{in}\right)^2\right)/\left(\delta S_2^{in}\right)^2$, as shown in the figure. $k^2$ is then used to establish the relation between canonical variables of light and canonical variables of memory and to find the variances and mean values of atomic canonical variables, as described in the main text.

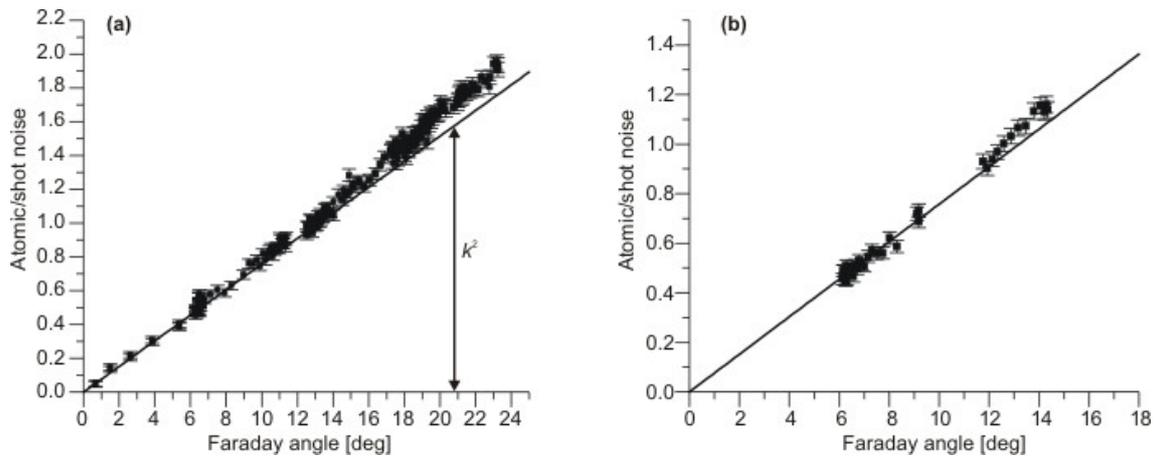

Supplementary Figure, The projection noise calibration. The atomic noise in units of the shot noise of light is plotted as a function of the macroscopic spin size $J_x$ which is proportional to the detected Faraday rotation angle. The error bars are statistical, arising from the fact that the noise variances are obtained from 10.000 cycles of the experiment. An increase in the noise level at high atomic densities seen in part **(a)** of the figure is due to the classical noise of the lasers. The solid line – the graph of $k^2$ - is the best estimate for the projection noise contribution. The value of $k^2$ for a particular experimental value of $J_x$ is shown with the arrow. In part **(b)** the PNL calibration experiment is repeated 43 days later and the same calibration still holds.

**Supplementary Notes**

**Derivation of the quantum feedback relations for an arbitrary quantum state of light**

Here we present a rigorous justification for the feedback relations used in the theoretical part of the paper. In the protocol we make a measurement on the operator of light $\hat{X}_L$ and then displace the atomic ensemble in momentum $\hat{P}_A$ by a quantity proportional to the outcome $x$. Denote the state of light and atoms after they have interacted as $|\Psi\rangle_{LA}$. Then, the non-normalized state after the measurement is $_L\langle x|\Psi\rangle_{LA}$. After the displacement the state is $\exp\{-ikx\hat{P}_A\}_L\langle x|\Psi\rangle_{LA}$. We can write this as $_L\langle x|\exp\{-ik\hat{X}_L\hat{P}_A\}|\Psi\rangle_{LA}$ by using the fact that $|x\rangle$ is a (generalized) eigenvalue of the operator $\hat{X}_L$. We now calculate the density operator obtained by averaging with respect to all outcomes of the measurement (with corresponding probability)

$$\rho = \int_{-\infty}^{\infty} dx \exp\{-ikx\hat{P}_A\}_L\langle x|\Psi\rangle_{LALA}\langle\Psi|x\rangle_L \exp\{ikx\hat{P}_A\} = \text{Tr}_m\left(\exp\{-ik\hat{X}_L\hat{P}_A\}|\Psi\rangle_{LALA}\langle\Psi|\exp\{ik\hat{X}_L\hat{P}_A\}\right)$$

where the trace is taken with respect to the measured mode. The averaged expectation value of any atomic operator $f(\hat{X}_A,\hat{P}_A)$ can be then determined by simply calculating its trace with this density operator. By using the cyclic property of the trace, we can re-express this quantity as the expectation value of the atomic operator in the Heisenberg picture which is obtained by displacing the atomic momentum operator by the light operator $\hat{X}_L$, i.e. $f(\hat{X}_A,\hat{P}_A+g\hat{X}_L)$. Thus, we can carry out the whole procedure in the Heisenberg picture by performing such a displacement. This is precisely what is done in the paper where the outcome of the measurement $x = \hat{X}_L^{in} + k\hat{P}_A^{in}$ is fed back into the atomic variable $\hat{P}_A$ with a feedback gain coefficient $g$. The result used in the paper is $\hat{P}_A^{mem} = \hat{P}_A^{in} - gx = \hat{P}_A^{in}(1-kg) - g\hat{X}_L^{in}$.

Note that this analysis is valid for arbitrary input states including mixed states.